\documentclass[onecollarge,natbib]{svjour2}
\bibpunct{[}{]}{;}{n}{}{,} 
\smartqed  
\usepackage{graphicx}
\usepackage{multirow}

 \usepackage{mathptmx}      
\usepackage{amsmath}
\usepackage{natbib}
\newcommand{\stm}{Skorniakov--Ter-Martirosian equation}

\journalname{Few body systems}
\begin{document}

\title{Universal physics of 2+1 particles with non-zero angular momentum\thanks{This work was supported by Grants-in-Aid (KAKENHI 22340114 and 22103005), the Global COE Program ``the Physical Sciences Frontier," and the Photon Frontier Network Program of MEXT of Japan.}}


\author{Shimpei Endo        \and
        Pascal Naidon    \and
        Masahito Ueda
}

\authorrunning{S. Endo, P. Naidon, and M. Ueda} 

\institute{S. Endo
          \at 
              Department of Physics, University of Tokyo, 7-3-1, Hongo, Bunkyo-ku, Tokyo, 113-0033, Japan \\
              Tel.: +81-3-5841-8326\\
              \email{endo@cat.phys.s.u-tokyo.ac.jp}           
             \and
             M. Ueda \and   P. Naidon \at
             Department of Physics, University of Tokyo, 7-3-1, Hongo, Bunkyo-ku, Tokyo, 113-0033, Japan \\
             ERATO Macroscopic Quantum Control Project, JST, 2-11-16 Yayoi, Bunkyo-ku, Tokyo 113-8656, Japan\\
}

\date{}

\maketitle

\begin{abstract}
The zero-energy universal properties of scattering between a particle and a dimer that involves an identical particle are investigated for arbitrary scattering angular momenta. For this purpose, we derive an integral equation that generalises the Skorniakov - Ter-Martirosian equation to the case of non-zero angular momentum. As the mass ratio between the particles is varied, we find various scattering resonances that can be attributed to the appearance of universal trimers and Efimov trimers at the collisional threshold. 
\end{abstract}
\section{Introduction}
\label{intro}
Few- and many-body systems at low energy with resonant short-range interactions are remarkable for their universal nature. In these systems, most of the details  of the interaction potentials are unimportant except for a few parameters, which fully characterize the interaction between the particles. In many cases, these parameters reduce to a single one, the $s$-wave scattering length $a$, and all the physical properties are expressed in terms of the scattering length $a$. For example, the binding energy of two particles scales as $a^{-2}$, and the scattering properties between these dimers and other particles or other dimers can also be universally expressed in terms of $a$ \cite{stm,PhysRevA.67.010703,petrov2004weakly}. When the $s$-wave scattering length becomes infinite, at some resonance point, one reaches the so-called unitarity regime, where the system has no intrinsic length scale and becomes completely universal \cite{ho2004universal}.



In some other cases, however, extra parameters are needed. This is the case of three-body systems where the Efimov effect occurs \cite{efimov1970energy,efimov1973energy}. The Efimov effect is characterized by the appearance of an infinite series of trimer bound states at unitarity. 
This effect results from the effective 3-body attraction  between particles  interacting through resonant 
short-range interactions. When this attraction is strong enough to overcome the repulsion that arises 
either from the centrifugal barrier due to rotation or from the Pauli exclusion principle of identical fermions, the 
particles can bind to form trimers called Efimov trimers. Because the particles are attracted to short 
distances, an extra short-range parameter is needed to completely determine their wave function. Thus, trimer 
binding energies and scattering properties in such systems depend on both the scattering length and the 
extra parameter. 
Efimov trimers are associated with remarkable features such as discrete scale invariance of their spectrum, and have attracted a lot of interest since their recent experimental realizations with ultracold atoms \cite{kraemer2006evidence,photoasso_Efimov_Lompe,nakajima2010measurement,PhysRevLett.103.130404,PhysRevLett.103.043201,pollack2009universality,gross2009observation}. 

In general, the strength of the Efimov attraction depends on the mass ratios between particles. For example,  in the case of two identical fermions and one different particle, which is resonantly interacting with the two fermions, the Efimov attraction is too weak for the Efimov effect to occur if the particles have the same masses. However, when the mass ratio between one of the two fermions and the third particle becomes larger than the critical value 13.6, and the system has one unit of total angular momentum, the Efimov attraction becomes dominant, and can support bound states, i.e. trimers formed of two fermions and one particle. 

Recently, Kartavtsev and Malykh found that trimers also exist for mass ratios smaller than the critical value \cite{kartavtsev2007low}. Indeed, even though in this regime the centrifugal repulsion dominates at short distance, its combined effect with the weak Efimov attraction creates a shallow potential well that can support these trimers. Although these trimers are related to the Efimov physics, their nature differs from that of Efimov trimers. First, they do not exhibit the discrete scale invariance of the Efimov trimer spectrum, and appear only for positive scattering lengths. Second, they are universal states in the sense that they can be characterized only by the scattering length, unlike Efimov states which also depend on short-range physics. For the universal trimers, the centrifugal barrier at short distance prevents the particles from coming close to each other, so that the three-body recombination occurring at a short distance is also suppressed. Thus, these universal trimer states are much longer-lived than the Efimov trimers \cite{levinsen2009atom}.

 In this paper we investigate the general problem of the particle-dimer scattering in the zero-range limit, where two of the particles are identical (bosonic or fermionic) particles and a third particle is of a different kind. In particular, we look for the resonances in the particle-dimer scattering due to the formation of the universal trimers. For this purpose, we derive an integral equation to calculate the particle-dimer scattering lengths in various angular momentum channels. This equation is a generalization of the \stm . With this integral equation, we calculate the particle-dimer scattering lengths for various angular momenta.We confirm that the scattering lengths diverge at the mass ratio where the universal trimers were predicted to appear  \cite{kartavtsev2008universal_JETP,kartavtsev2008universal_FBS}.

The paper is organized as follows. In Sec. \ref{sec:1}, a generalization of the \stm \ to an arbitrary-angular momentum is presented.  In Sec. \ref{sec:2}, we show the results of the calculation of the particle-dimer scattering lengths for various angular momenta, and the binding energy of the universal trimer states.  In the Appendix, we give the detailed derivation of the generalized Skorniakov--Ter-Martirosian equation for high angular momenta.

\section{Generalized Skorniakov--Ter-Martirosian integral equation for arbitrary angular momenta}
\label{sec:1}

We consider a system of two identical particles (bosons or fermions) strongly interacting with a third particle of a different kind.  We assume that this interaction is nearly resonant, and thus characterized by an $s$-wave scattering length $a$, which is much larger than the range of the interaction. On the other hand, we assume that the interaction between the identical particles is not resonant, and therefore can be neglected. This is even more justified for identical fermions, which do not interact in the $s$-wave channel. As a result, there is only one length scale in the problem, given by $a$, which we take as the unit length: $a=1$.

This three-body problem can be solved by an integral-equation approach. The $s$-wave scattering length between the particle and the dimer, in which one of the particles is identical with the unbound one, can be obtained by solving the integral equation derived by  Skorniakov and Ter-Martirosian \cite{stm}, and found to be $1.18$ in the equal-mass case. Recently, the calculation was also extended to the unequal-mass case \cite{PhysRevA.67.010703}. Here, we derive a similar integral equation to calculate the scattering lengths and the binding energy for arbitrary angular momenta. 

The quantity characterizing the particle-dimer scattering in the $l$-th wave angular momentum channel is the $l$-th wave scattering lengths $a_l$, which are defined as a low-energy contribution of the $l$-th wave phase shift $\delta_l$
\begin{equation}
k^{2l+1}\cot \delta_l = -\frac{1}{a_l}+O(k^2).
\end{equation}
Since the phase shift $\delta_l$ is related with the $l$-th wave angular momentum component of the particle-dimer scattering amplitude $f_l(k)$ as
\begin{equation}
f_l(k)=\frac{1}{k\cot \delta_l - ik},
\end{equation}
$a_l$ represents the low-energy scattering property \cite{landau1991quantum}. The partial wave scattering length $a_l$ can be calculated from the following integral equation, whose derivation is given in the Appendix:
\begin{equation}
\begin{split}
\label{eq:stm}&\frac{(-1)^{l+1} 2^{2l+1}(l!)^2}{(2l+1)!}\frac{\alpha^l}{(2\alpha+1)(\alpha+1)^{l-2}}\Bigr[\frac{1}{(-\epsilon +p_1^2)}\Bigl]^{l+1}\\
=&\pm a_l(p_1) +\frac{(-1)^{l+1}}{\pi}\frac{(\alpha+1)}{\alpha}\int_{0}^{\infty} dq \frac{q^{l+1}}{p_1^{l+1}}Q_l\Biggr(\frac{\frac{\alpha+1}{{2\alpha}}[-\epsilon +p_1^2+q^2]}{p_1q}\Biggl)\frac{1}{\sqrt{-\epsilon+\frac{2\alpha+1}{(\alpha+1)^2}q^2}-1}a_l(q),
\end{split}
\end{equation}
where $\pm$ refers to bosons (+) and fermions (-), $\alpha$ is the mass ratio between one of the two identical particles and the third particle, $\epsilon$ is the total energy in units of the dimer's binding energy, $Q_l$ denotes the Legendre function of the second kind, and $a_l(q)$ is the $l$-th wave momentum-dependent scattering length whose value at $q=0$ corresponds to the $l$-th wave scattering length $a_l$. By solving this integral equation at the dimer-particle continuum threshold $\epsilon=-1$, we obtain the $l$-th wave scattering length. On the other hand, we can find the binding energy of the trimer states by searching for the value of $\epsilon$ at which $a_l$ diverges. This can be done by finding the value of $\epsilon$ at which the eigenvalue of the right-hand side (seen as an operator acting on $a_l(q)$) vanishes.

We note that in the numerical calculation, we take a finite momentum cutoff instead of the infinite upper boundary of the integral in Eq. (\ref{eq:stm}). For a mass ratio smaller than the Efimov threshold listed in Table \ref{tab:result_list}, we obtain convergent results if we take a sufficiently large momentum cutoff. However, if the mass ratio gets closer to the Efimov threshold, the cutoff required for the convergence to the universal results becomes larger and larger. Above the Efimov threshold, the results depend on the cutoff, as it plays the role of the short-range parameter needed to characterize Efimov physics \cite{naidon2010efimov}.

\begin{figure*}
\centering
\includegraphics[width=0.9\textwidth]{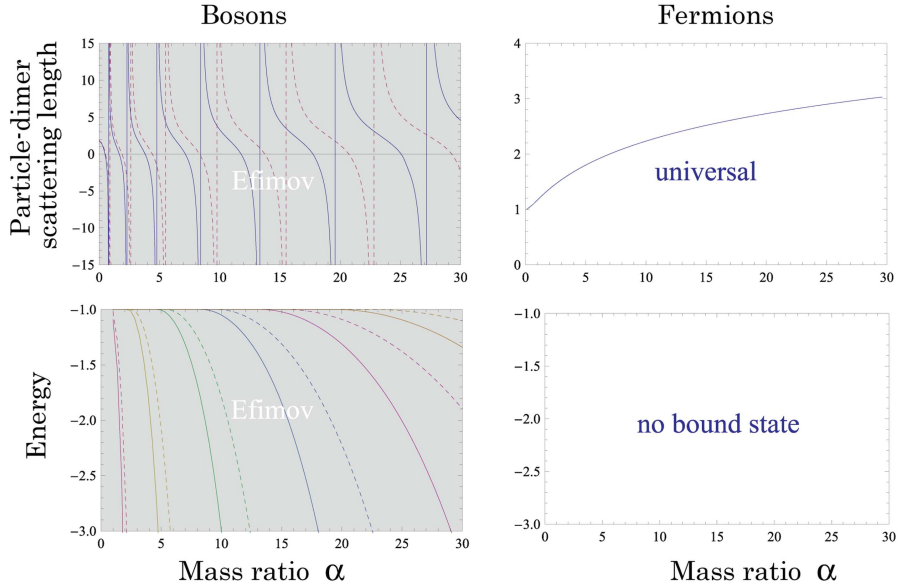}
\caption{$S$-wave scattering length and binding energy for the bosonic and fermionic cases. (a),(b) $s$-wave scattering length, and (c), (d) binding energy of trimers are shown. The particle-dimer scattering length is displayed in units of the particle-particle scattering length $a$, while the energy is shown in units of the dimer binding energy. The gray and white regions denote the universal and non-universal regions, respectively. The dashed curves correspond to a momentum cutoff of $k_c=2000$ and the solid curve $k_c=4000$.}
\label{fig:1}       
\end{figure*}

\begin{figure*}
\centering
\includegraphics[width=0.9\textwidth]{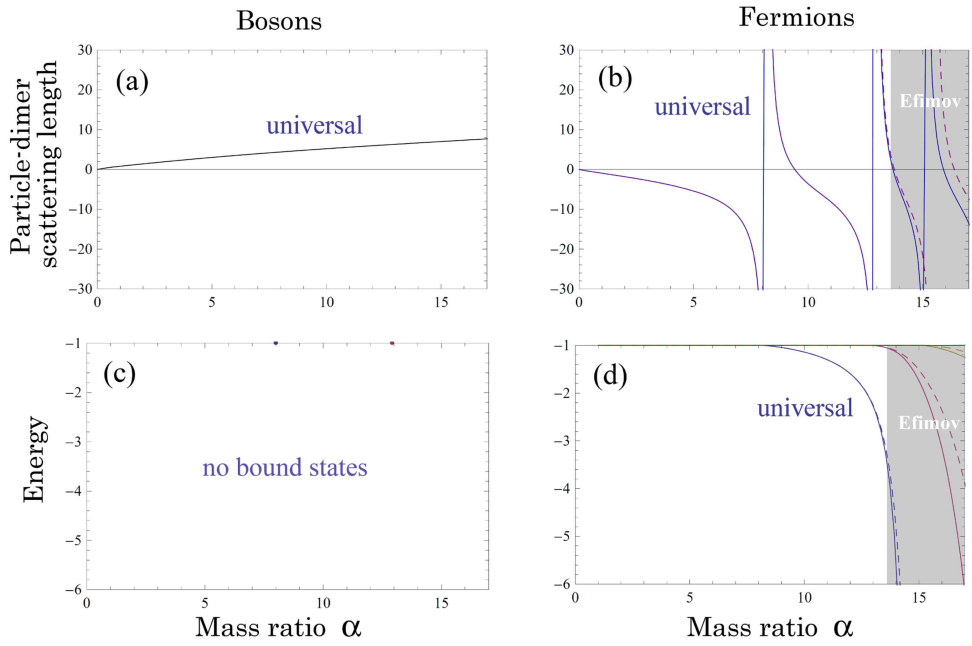}
\caption{$P$-wave scattering volume and binding energy. Notations are the same as those in Fig. \ref{fig:1}.}
\label{fig:2}       
\end{figure*}

\begin{figure*}
\centering
\includegraphics[width=0.9\textwidth]{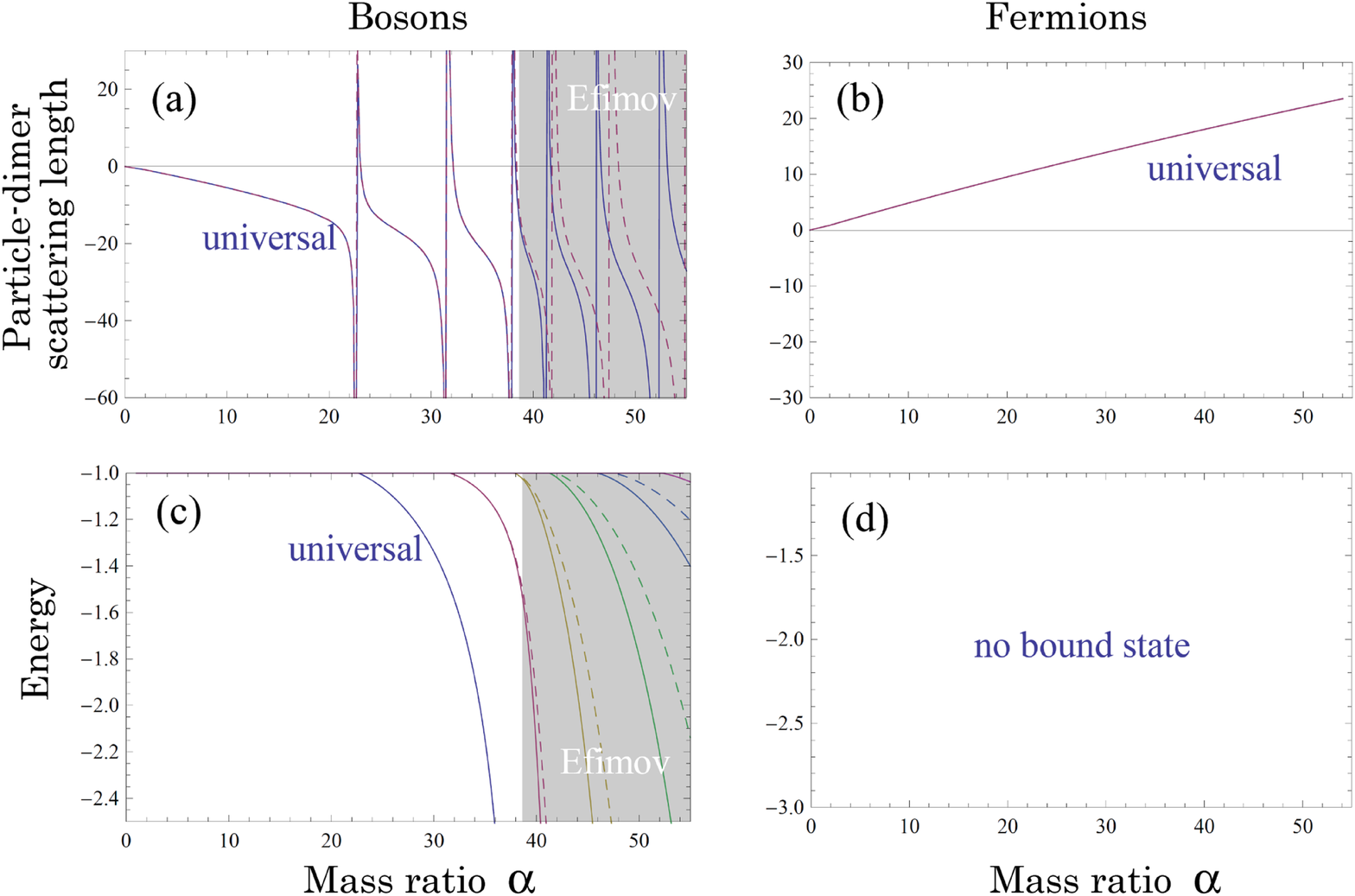}
\caption{$D$-wave scattering length and binding energy for the bosonic and fermionic cases. Notations are the same as those in Fig. \ref{fig:1}.}
\label{fig:3}       
\end{figure*}

\begin{figure*}
\centering
\includegraphics[width=0.9\textwidth]{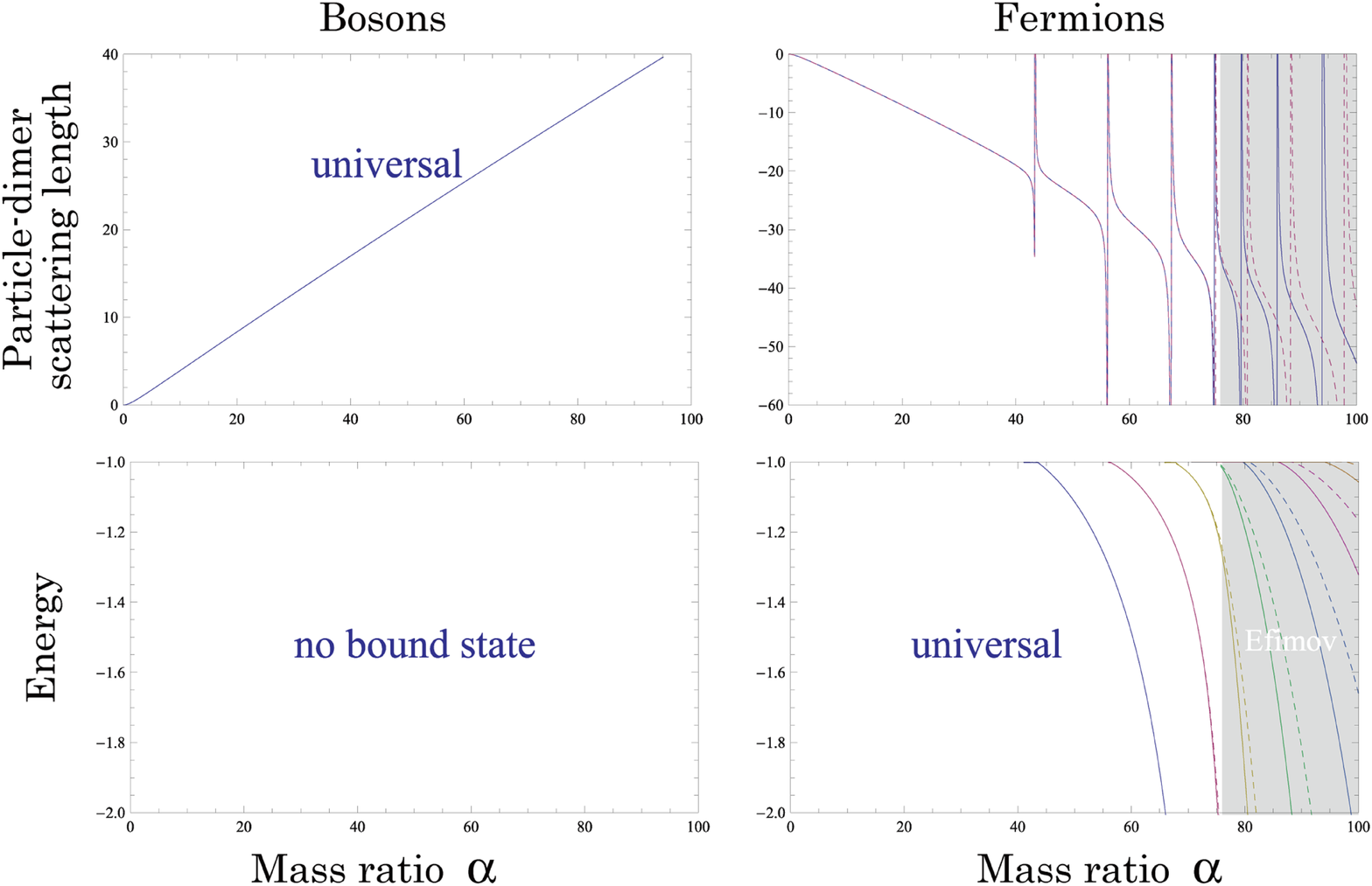}
\caption{$F$-wave scattering length and binding energy for the bosonic and fermionic cases. Notations are the same as those in Fig. \ref{fig:1}.}
\label{fig:4}       
\end{figure*}

\begin{figure*}
\centering
\includegraphics[width=0.9\textwidth]{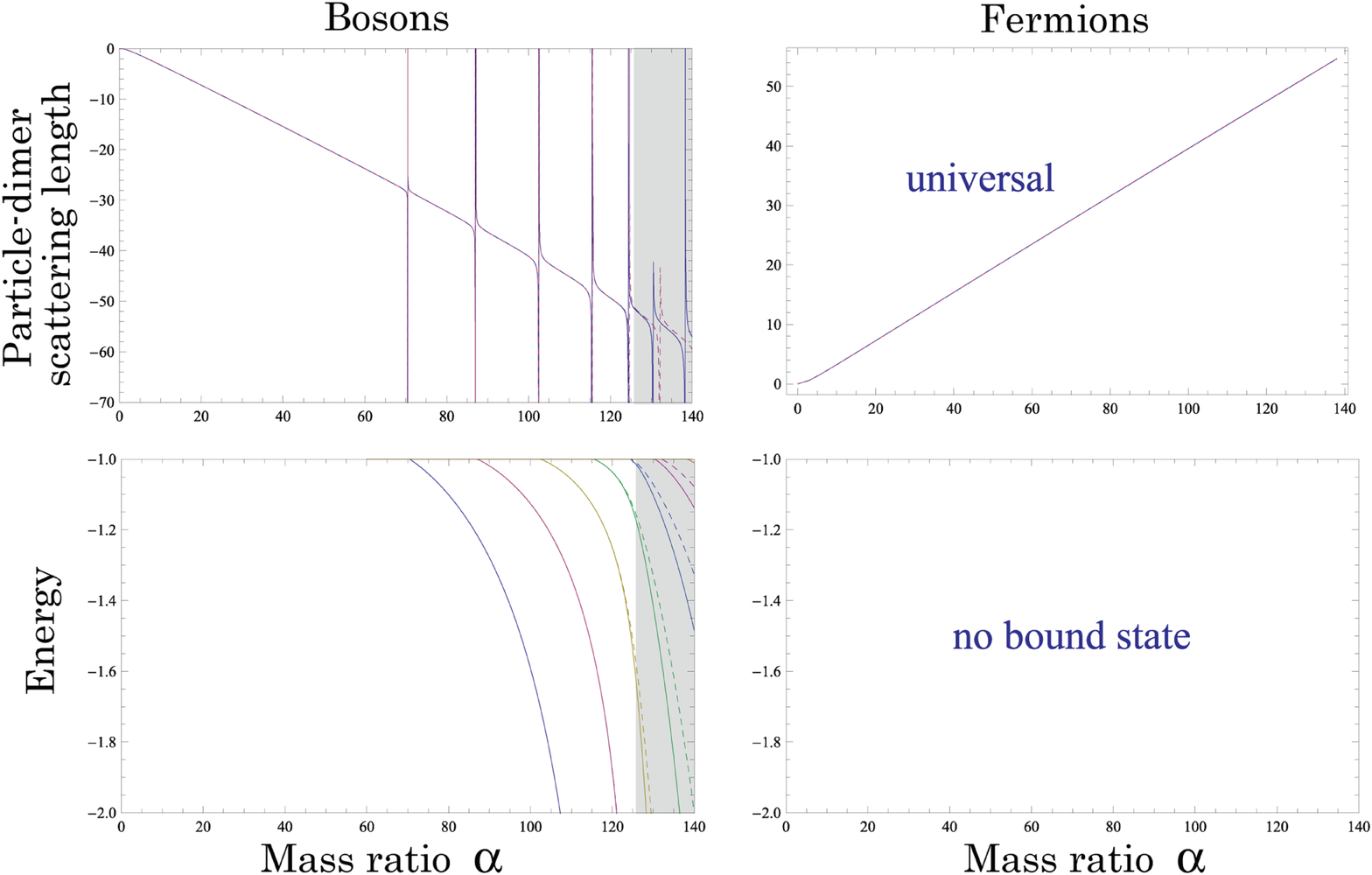}
\caption{$G$-wave scattering length and binding energy for the bosonic and fermionic cases. Notations are the same as those in Fig. \ref{fig:1}.}
\label{fig:5}       
\end{figure*}

We display the results for both bosons and fermions in Fig. \ref{fig:1} to Fig. \ref{fig:5}. 
For the calculations performed at a mass ratio smaller than the Efimov threshold given in Table \ref{tab:result_list}, the results should not depend on the momentum cutoff in the integration if we take it as sufficiently large. This is illustrated in the figures by a set of curves corresponding to different momentum cutoffs. Below the Efimov threshold, these curves essentially superimpose and correspond to the universal curve. Above the Efimov threshold, there is no universal curve and the results explicitly depend on the cutoff in this region.


\begin{table}
\caption{Mass ratio at which the universal bound state starts to appear, the width of the particle-dimer resonance, and the Efimov threshold are shown for various angular momenta and statistics. For the case of $l=1$, for example, the universal bound states appear at the mass ratios of 8.172 and 12.917, which reproduces the results of Ref. \cite{kartavtsev2007low}. For higher-angular momenta, we refine the values of Ref. \cite{kartavtsev2008universal_JETP}. We also present the Efimov thresholds, above which the non-universality appears and the Efimov states can be formed \cite{efimov1973energy,nielsen2001three}.}
\centering
\label{tab:result_list} 
\begin{tabular}{ccccc}
\hline\noalign{\smallskip}
$l$ & statistics & universal bound state & width of the resonance & Efimov threshold\\[3pt]
\tableheadseprule\noalign{\smallskip}
\multirow{2}{*}{$1$} & \multirow{2}{*}{fermions} & $8.172$   &   $9.2$ &\multirow{2}{*}{$13.607$}\\
        &                &  $12.917$ &  $8.1$  & \\
\hline
\multirow{3}{*}{$2$} & \multirow{3}{*}{bosons} & $22.637$  & $5.9$ & \multirow{3}{*}{$38.630$}\\
     &                 &  $31.523$ &  $11.9$ & \\
       &                 &  $37.766$ &  $8.1$ & \\
\hline
\multirow{3}{*}{$3$} & \multirow{3}{*}{fermions} &  $43.395$ &  $1.3$ & \multirow{3}{*}{$75.994$}\\
       &                 &  $56.166$ &  $3.7$ & \\
   &                 &  $67.336$ &  $5.4$ & \\
\hline
\multirow{5}{*}{$4$} & \multirow{5}{*}{bosons} &   $70.457$ & $ 0.14 $& \multirow{5}{*}{$ 125.765$}\\
    &   &   $87.027$ & $ 0.50 $& \\
   &   &   $115.534$ & $ 0.99 $&  \\
   &   &   $102.486$ & $ 1.27 $&  \\
      &   &   $124.167$ & $ 0.81 $&  \\
\noalign{\smallskip}\hline
\end{tabular}
\end{table}

\section{Results of numerical calculations}
\label{sec:2}

In the case of $s$-wave and bosons (Fig. \ref{fig:1} (a) and (c)), we have an infinite series of resonances due to the formation of bound states. Because there is no centrifugal repulsion and the Efimov attraction prevails for all mass ratios, all these bound states are Efimov states, and therefore depend on the cutoff of the integral. As we change the cutoff, this series of resonances and bound states gradually shift, while preserving their qualitative structure imposed by discrete scale invariance.


For the $s$-wave fermionic case (Fig. \ref{fig:1} (b) and (d)), on the other hand, Pauli's exclusion principle suppresses Efimov's attraction, and the results are universal and do not depend on the cutoff. The scattering length curve presented here reproduces the results already presented in the previous studies \cite{stm,PhysRevA.67.010703,Iskin_uneq_mass}. 

For the $p$-wave fermionic case (Fig. \ref{fig:2} (b) and (d)), we can see two universal $p$-wave resonances followed by other non-universal $p$-wave resonances. The two universal resonances are due to the formation of the bound states originally found by Kartavtsev and Malykh \cite{kartavtsev2007low}. The non-universal resonances occur at mass ratios larger than the Efimov critical value 13.6, and are thus due to a series of Efimov states. For $p$-wave bosonic case (Fig. \ref{fig:3}), on the other hand,  the effective potential is repulsive due to the symmetry of the identical particles, and thus the curve is universal and there is no bound state. The widths of the resonances in the fermionic case are broad, suggesting that the universal resonances may be observed even when the mass ratio is not finely tuned.

In Fig. \ref{fig:3} to  Fig. \ref{fig:5}, we present the scattering lengths and binding energy in the higher angular-momentum channels. We can clearly see that there are several universal bound states for bosons in the even angular-momentum channels, and for fermions in the odd angular-momentum channels.  This reconfirms the conclusion of the previous works \cite{kartavtsev2008universal_JETP,kartavtsev2008universal_FBS}. The mass ratio at which the resonances occur also agrees with the value where the universal trimers were predicted to appear \cite{kartavtsev2008universal_JETP,kartavtsev2008universal_FBS}. By looking at these binding energy curves, it becomes apparent that they can be grouped in two sets. One corresponds to the Efimov series, where states are related to each other by a scale transformation. The other one corresponds to the universal trimer states, which were shown in Ref. \cite{kartavtsev2008universal_JETP} to be related to a universal function of only two parameters. We note that the widths of the resonances become narrower in the higher angular-momentum channels.




\section{Conclusion}
We have studied the particle-dimer scattering for the unequal-mass case for both bosons and fermions. For this purpose, we have derived an integral equation for arbitrary angular-momentum channels, which is a generalization of the \stm. With this integral equation, we have
calculated the particle-dimer scattering lengths for various angular-momentum channels, and found resonances due to the formation of universal trimers and Efimov trimers.

Ultracold atomic gases are currently the most likely candidates for the experimental possibility of observing the universal resonances and bound states. In these systems, a very large $s$-wave scattering length can be realized by using a Feshbach resonance, providing access to the universal regime where the scattering length is the single parameter characterizing the system \cite{inouye1998observation}. As for controlling the mass ratio, it is possible to prepare hetero-nuclear systems with mass imbalance \cite{PhysRevLett.103.043201}, and thus select the appropriate mass ratio from a broad variety of atomic species. Some possible atomic combinations are listed in Table \ref{tab:species}. We also note that several works have also shown that reduced dimensionality or strong confinement favours the appearance of similar resonances  and trimer states at lower mass ratios \cite{levinsen2009atom,kartavtsev2009bound_JETP,PhysRevA.82.033625}

\begin{table}
\caption{Mass ratios for some of the atomic hetero-nuclear systems}
\centering
\label{tab:species}       
\begin{tabular}{ccc}
\hline\noalign{\smallskip}
species & statistics & mass ratio  \\
\hline
$^7$Li-$^{87}$Sr-$^{87}$Sr & fermions & 12.4 \\
$^6$Li-$^{87}$Sr-$^{87}$Sr & fermions & 14.4 \\
$^7$Li-$^{133}$Cs-$^{133}$Cs & bosons & 18.9 \\
$^6$Li-$^{133}$Cs-$^{133}$Cs & bosons & 22.1 \\
$^7$Li-$^{174}$Yb-$^{174}$Yb & bosons & 24.8 \\
$^6$Li-$^{174}$Yb-$^{174}$Yb & bosons & 28.9 \\
\hline
\end{tabular}
\end{table}

If these resonances can be achieved experimentally with ultracold gases, they should provide better control over the scattering property than the Efimov resonances. Because atomic Efimov states are inherently unstable due to a recombination to lower vibrational states, the atom-dimer Efimov resonances are dominated by the strong inelastic collisions and smoothed out. On the other hand, the centrifugal barrier of the universal trimers prevents them from undergoing such recombination processes, and makes them stable \cite{levinsen2009atom}. Thus, they are promising for controlling interactions in ultracold gases. In particular, the comparatively broad $p$-wave resonances seem to be most feasible.




\begin{acknowledgements}
We thank D. Petrov for introducing us to the work of Ref. \cite{kartavtsev2007low}. We also thank N. Sakumichi and Y. Tanizaki for fruitful discussions. 
\end{acknowledgements}

%
%

\appendix
\section{Derivation of the higher-angular-momentum \stm}
\label{sec:higher_stm}
We here derive an equation to determine the higher angular-momentum scattering length denoted as $a_l$. We start from the three-body Dyson's equation \cite{PhysRevA.73.032724}
\begin{equation}
\label{eq:three_Dyson}T_3(p_1,p_2|P) = \pm G_{t}(P-p_1-p_2) \pm  i\int\frac{d\omega_q d\vec{q}}{(2\pi)^4 }G_{t}(P-p_1-q)G_{i}(q)T_2(P-q)T_3(q,p_2|P),
\end{equation}
where $T_2(q)$ and $T_3(p_1,p_2|P)$ are the 2-body and 3-body $T$-matrices in the vacuum, and $G_{i}(q)$ is the single particle Green's functions in the vacuum ($s=1,2$) for identical particles with mass $m_i$, and $G_{i}(q)$ for the third particle with mass $m_t$. The plus sign is for bosons, and the minus sign for fermions. Note that all momentum indices are written in four-momentum notation.

We focus on the on-shell process for $p_2$ and collision at the total energy $P=P_0\equiv \{\frac{\varepsilon}{2\mu},\vec{0}\}$, where $\mu=\frac{m_i m_t}{m_i+m_t}$ is the reduced mass between the different particles, and $\epsilon$ is the total energy in units of the dimer's binding energy. Then, Eq. (\ref{eq:three_Dyson}) reduces to
\begin{equation}
\label{eq:tmp_32}T_3(\overline{p_1}^{i},\overline{p_2}^{i}|P_0) = \pm G_{t}(P_0-\overline{p_1}^{i}-\overline{p_2}^{i}) \pm \int \frac{d\vec{q}}{(2\pi)^3} G_{t}(P-\overline{p_1}^{i}-\overline{q}^{i})T_2(P_0-\overline{q}^{i})T_3(\overline{q}^{i},\overline{p_2}^{i}|P_0),
\end{equation}
where the overline stands for the on-shell value for the one of the identical particles: $\overline{q}^{i}\equiv \{\frac{\vec{q}^2}{2m_{i}}, \vec{q}\}$. Our main motivation is to expand this equation into partial waves. To this end, we will first focus on the terms such as
\begin{equation}
-G_{t}(P-\overline{p_1}^{i}-\overline{p_2}^{i})=\frac{1}{-\frac{\varepsilon}{2\mu}+\frac{\vec{p}_1^2}{2m_{i}}+\frac{\vec{p}_2^2}{2m_{i}}+\frac{(\vec{p}_1+\vec{p}_2)^2}{2m_{t}}}=\frac{1}{-\frac{\varepsilon}{2\mu}+\frac{\vec{p}_1^2}{2\mu}+\frac{\vec{p}_2^2}{2\mu}+\frac{\vec{p}_1\vec{p}_2}{m_{t}}},
\end{equation}
which couples two ingoing and outgoing momentum. If this term can be expanded as
\begin{equation}
-G_{t}(P_0-\overline{p_1}^{i}-\overline{p_2}^{i})=\sum_{l=0}^{\infty}g_l(p_1,p_2)P_l(\cos\theta_{12}),
\end{equation}
where $\theta_{12}$ is the relative angle between $\vec{p_1}$ and  $\vec{p_2}$, then we can easily expand Eq. (\ref{eq:tmp_32}). In fact, if we expand the $T$-matrix as
\begin{equation}
T_3(\overline{q}^{i},\overline{p_2}^{i}|P_0)=\sum_{l=0}^{\infty}t_l(p_1,p_2)P_l(\cos\theta_{12})
\end{equation}
and substitute this into Eq. (\ref{eq:tmp_32}), we obtain
\begin{equation}
\label{eq:t_l_eq}t_l(p_1,p_2)=\mp g_l(p_1,p_2) \mp \frac{1}{2\pi^2(2l+1)}\int_{0}^{\infty} q^2dq g_l(p_1,q)T_2(P_0-\overline{q}^{i})t_l(q,p_2),
\end{equation}
where we have used the following formula:
\begin{equation}
P_l(\cos\theta_{12})=\frac{4\pi}{2l+1}\sum_{m=-l}^{l} Y_{lm}(\theta_1,\phi_1)Y_{lm}^{*}(\theta_2,\phi_2).
\end{equation}
Thus, we only have to solve the above integral equation to obtain the $l$-th partial wave scattering length. For this purpose we have to calculate $g_l(p_1,p_2)$. From the orthonormal relation of Legendre polynomials
\begin{equation}
\int_{-1}^{1}dx P_n(x)P_m(x)=\frac{2}{2n+1}\delta_{n,m},
\end{equation}
we can calculate $g_l(p_1,p_2)$ from the following integration
\begin{equation}
g_l(p_1,p_2)=\frac{2l+1}{2}\int_{-1}^{1}dx P_l(x)\frac{1}{-\frac{\varepsilon}{2\mu}+\frac{p_1^2}{2\mu}+\frac{p_2^2}{2\mu}+\frac{p_1p_2}{m_{t}}x}.
\end{equation}
Here we use the following relation, which we will show below:
\begin{equation}
\label{eq:legendre_integral}\int_{-1}^{1}dx P_n(x)\frac{1}{1+Ax}=  2(-1)^n \frac{1}{A}Q_n\Bigr(\frac{1}{A}\Bigl)\ \ (0<A<1) ,
\end{equation}
where $Q_n(x)$ is a Legendre function of the second kind. Then, 
\begin{equation}
\label{eq:g_l_expression}g_l(p_1,p_2)=(2l+1)(-1)^l \frac{m_{t}}{p_1p_2}Q_l\Biggr(\frac{\frac{m_{t}}{{2\mu}}[-\varepsilon +p_1^2+p_2^2]}{p_1p_2}\Biggl).
\end{equation}
\begin{figure}
\centering
\includegraphics[width=9cm,clip,angle =0]{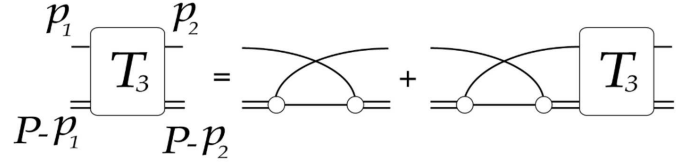}
\caption{Diagrammatic representation of three-body Dyson's equation (\ref{eq:three_Dyson}). The single curves and the double curves denote single-particle Green's functions and two-body $T$-matrices, respectively.}
\label{fig:3body_Dyson}
\end{figure}
We are only interested in the low-energy value of $t_l(p_1,p_2)$ under $p_1=p_2$, since we are interested in the low-energy elastic collision. In fact, the $T$-matrix and the scattering amplitude are related as
\begin{equation}
T_3(\overline{p_1}^{i},\overline{p_2}^{i}|P_0)=-\frac{\mu^2}{m_{DA}}f(\vec{p_1},\vec{p_2}),
\end{equation}
where $m_{DA}=\frac{m_{i}(m_{i}+m_{t})}{2m_{i}+m_{t}}$ is the reduced mass between the particle and the dimer. This scattering amplitude and the $l$-th-wave scattering length are related as
\begin{eqnarray}
f(\vec{p_1},\vec{p_2})&=&\sum_{l}(2l+1)f_l(p_1)P_l(\cos\theta_{12}) \ (|\vec{p}_1|=|\vec{p}_2|),\\
f_l(k)&=&\frac{k^{2l}}{-\frac{1}{a_l}+r_l k^2+\cdots}.
\end{eqnarray}
Thus, we can relate the $l$-th-wave scattering length and the $T$-matrix as follows:
\begin{equation}
a_l =-\lim_{p_1 \rightarrow 0} \frac{f_l(p_1)}{p_1^{2l}}= \frac{1}{2l+1} \frac{m_{DA}}{\mu^2} \lim_{p_1,p_2\rightarrow 0}\frac{t_l(p_1,p_2)}{p_1^l p_2^l}.
\end{equation}
Thus, if we define a momentum-dependent $l$-th wave scattering length as
\begin{equation}
a_l(p_1)\equiv  \frac{1}{2l+1} \frac{m_{DA}}{\mu^2}\lim_{p_2\rightarrow 0}\frac{t_l(p_1,p_2)}{p_1^l p_2^{l}}.
\end{equation}
Then we can rewrite Eq. (\ref{eq:t_l_eq}) as
\begin{equation}
\label{eq:tmp_33}a_l(p_1)= \mp \frac{1}{2l+1} \frac{m_{DA}}{\mu^2}\frac{g_l(p_1)}{p_1^l} \mp  \frac{1}{2\pi^2(2l+1)p_1^l}\int_{0}^{\infty} dqq^{l+2} g_l(p_1,q)T_2(P_0-\overline{q}^{i})a_l(q),
\end{equation}
where we have defined $g_l(p_1)$ as
\begin{equation}
g_l(p_1)\equiv \lim_{p_2\rightarrow 0}\frac{g_l(p_1,p_2)}{p_2^{l}}.
\end{equation}
By using the following asymptotic expression for the Legendre function of the second kind
\begin{equation}
Q_n(z)\rightarrow \frac{2^n(n!)^2}{z^{n+1}(2n+1)!} \ (z\rightarrow \infty),
\end{equation}
we can show from Eq. (\ref{eq:g_l_expression}) that
\begin{equation}
\label{eq:g_l_expression_2}g_n(p_1)=\frac{(-2)^n(n!)^2 m_{t} p_1^n}{(2n)!} \Biggr(\frac{1}{\frac{m_{t}}{{2\mu}}[-\varepsilon +p_1^2]}\Biggl)^{n+1}.
\end{equation}
Using Eqs. (\ref{eq:g_l_expression}), (\ref{eq:tmp_33}), and (\ref{eq:g_l_expression_2}), we obtain
\begin{equation}
a_l(p_1)=\mp \frac{(-2)^{l}(l!)^2}{(2l+1)!}\frac{m_{DA}m_{t}}{\mu^2}\Bigr[\frac{2\mu}{m_{t}}\frac{1}{(-\varepsilon +p_1^2)}\Bigl]^{l+1} \pm \frac{(-1)^{l+1} m_{t}}{2\pi^2}\int_{0}^{\infty} dq\Bigr(\frac{q}{p_1}\Bigl)^{l+1}Q_l\Biggr(\frac{\frac{m_{t}}{{2\mu}}[-\varepsilon +p_1^2+q^2]}{p_1q}\Biggl)T_2(P_0-\overline{q}^{i})a_l(q).
\end{equation}
If we rewrite this in terms of mass ratio parameter $\alpha = m_{i}/ m_{t}$, we obtain
\begin{equation}
\begin{split}
\label{eq:appendix_last_mae}a_l(p_1)=&\pm \frac{(-1)^{l+1} 2^{2l+1}(l!)^2}{(2l+1)!}\frac{\alpha^l}{(2\alpha+1)(\alpha+1)^{l-2}}\Bigr[\frac{1}{(-\varepsilon +p_1^2)}\Bigl]^{l+1} \\
&\pm \frac{(-1)^{l+1} m_{t}}{2\pi^2}\int_{0}^{\infty} dq\Bigr(\frac{q}{p_1}\Bigl)^{l+1}Q_l\Biggr(\frac{\frac{\alpha+1}{{2\alpha}}[-\varepsilon +p_1^2+q^2]}{p_1q}\Biggl) T_2(P_0-\overline{q}^{i})a_l(q).
\end{split}
\end{equation}
We will use the expression 
\begin{equation}
\label{eq:t2_result}T_2(P_0-\overline{q}^{i}) = -\frac{2\pi}{\mu}\frac{1}{\sqrt{-\varepsilon+\frac{2\alpha+1}{(\alpha+1)^2}q^2}-1}.
\end{equation}
 Then, from Eq. (\ref{eq:appendix_last_mae}), we obtain
\begin{equation}
\begin{split}
a_l(p_1)=&\pm \frac{(-1)^{l+1} 2^{2l+1}(l!)^2}{(2l+1)!}\frac{\alpha^l}{(2\alpha+1)(\alpha+1)^{l-2}}\Bigr[\frac{1}{(-\varepsilon +p_1^2)}\Bigl]^{l+1} \\
&\pm \frac{(-1)^{l}}{\pi}\frac{(\alpha+1)}{\alpha}\int_{0}^{\infty} dq \frac{q^{l+1}}{p_1^{l+1}}Q_l\Biggr(\frac{\frac{\alpha+1}{{2\alpha}}[-\varepsilon +p_1^2+q^2]}{p_1q}\Biggl)\frac{1}{\sqrt{-\varepsilon+\frac{2\alpha+1}{(\alpha+1)^2}q^2}-1}a_l(q).
\end{split}
\end{equation}

Finally, we prove Eq. (\ref{eq:legendre_integral}).
We first use the following formula:
\begin{equation}
\frac{1}{a}= \int_{0}^{\infty} dt \exp(- a t) \ (a>0).
\end{equation}
Then, the left-hand side (LHS) of Eq. (\ref{eq:legendre_integral}) can be rewritten as
\begin{eqnarray}
\mathrm{LHS}&=& \int_{-1}^{1}dx\int_{0}^{\infty} dtP_l(x)\exp(- [1+Ax] t)\\
&=& 2(-1)^n \int_{0}^{\infty} dt e^{-t} i_n(At)\\
&=& 2(-1)^n \frac{1}{A}Q_n\Bigr(\frac{1}{A}\Bigl),
\end{eqnarray}
where $i_l$ is a spherical modified Bessel function and $Q_l$ is the Legendre function of the second kind. We have used the following formula for these functions in the second line and the third line:  
\begin{eqnarray}
\int_{-1}^{1}dxP_n(x)\exp(- b x)&=& 2(-1)^n i_n(b)\ (b>0),\\
\int_{0}^{\infty} dt i_{n}(t) e^{-pt}&=&Q_n(p)\ (p>0).
\end{eqnarray}
\qed


%
%

\end{document}